\documentclass[prl,reprint,amsmath,amssymb,preprintnumbers,superscriptaddress,nofootinbib]{revtex4-1}

\usepackage{epsfig,epstopdf}
\usepackage{subfigure}
\usepackage{color}

\def\beq{\begin{equation}}
\def\eeq{\end{equation}}
\def\bea{\begin{eqnarray}}
\def\eea{\end{eqnarray}}

\def\mev{{\rm MeV}}

\def\eps{\varepsilon}

\newcommand{\lsim}{
\mathrel{\hbox{\rlap{\hbox{\lower4pt\hbox{$\sim$}}}\hbox{$<$}}}}
\newcommand{\gsim}{
\mathrel{\hbox{\rlap{\hbox{\lower4pt\hbox{$\sim$}}}\hbox{$>$}}}}

\def\eps{\varepsilon}

\begin{document}

\preprint{CTPU-16-34}
\title{Portal Connecting Dark Photons and Axions}
%\title{Dark Axion Portal}

\author{Kunio Kaneta}
\affiliation{Center for Theoretical Physics of the Universe, Institute for Basic Science (IBS), Daejeon 34051, Korea}
\author{Hye-Sung Lee}
\affiliation{Center for Theoretical Physics of the Universe, Institute for Basic Science (IBS), Daejeon 34051, Korea}
\author{Seokhoon Yun}
\affiliation{Center for Theoretical Physics of the Universe, Institute for Basic Science (IBS), Daejeon 34051, Korea}
\affiliation{Department of Physics, KAIST, Daejeon 34141, Korea}

\date{November 2016}

\begin{abstract}
\noindent 
The dark photon and the axion (or axion-like particle) are popular light particles of the hidden sector.
Each of them has been actively searched for through the couplings called the vector portal and the axion portal.
We introduce a new portal connecting the dark photon and the axion (axion--photon--dark photon, axion--dark photon--dark photon), which emerges in the presence of the two particles.
This dark axion portal is genuinely new couplings, not just from a product of the vector portal and the axion portal, because of the internal structure of these couplings.
We present a simple model that realizes the dark axion portal and discuss why it warrants a rich phenomenology.
\end{abstract}

\maketitle

%%%%%%%%%%%%%%%%%%%%%%%%%%%%
{\em Introduction.---}
%%%%%%%%%%%%%%%%%%%%%%%%%%%%
Despite the success of the standard model (SM), there are reasons to believe that our physical world is composed of more particles, including those that do not couple to the SM particles in a vivid way (e.g., dark matter); these are often called the hidden sector particles.

Because of the small couplings to the SM particles, a hidden sector particle can be much lighter than the electroweak scale.
The concept of {\em portals} has helped in understanding the possible mixing of the hidden sector particles with the SM and how to search for them.
Known portals include \\
$~~~~~~~~~~$ (i) Vector portal:  $B_{\mu\nu} Z'^{\mu\nu}$, \\
$~~~~~~~~~~~$(ii) Axion portal:  $(a / f_a) F_{\mu\nu} \tilde F^{\mu\nu} , \cdots$, \\
$~~~~~~~~~~~$(iii) Higgs portal: $|S|^2 H^\dagger H , \cdots$, \\
$~~~~~~~~~~~$(iv) Neutrino portal: $LHN$. \\
For a review on the relevant physics of these portals, see Ref.~\cite{Essig:2013lka}.
The relic dark matter (DM) can be either a portal particle or just coupled to a portal particle via a hidden interaction.
For instance, it may couple to a new vector boson \cite{ArkaniHamed:2008qn} or an axion-like particle (ALP) \cite{Nomura:2008ru}.

There are natural setups that can combine the portals.
For example, the vector portal and Higgs portal can coexist because a Higgs singlet can provide a mass to a new vector boson (or dark photon).
It then suggests a new search scheme of the hidden sector particles, for instance, a rare Higgs decay, $H \to Z' Z' \to$ four-lepton state where the first decay relies on the Higgs portal and the second decay relies on the vector portal \cite{Gopalakrishna:2008dv}.
This is basically based on the product of the two portals.
(For more recent studies on this and the related topics, see Refs.~\cite{Davoudiasl:2012ag,Davoudiasl:2012ig,Davoudiasl:2013aya,Curtin:2013fra,Curtin:2014cca}.)

\begin{figure}[t]
\centering
\includegraphics[width=0.47\textwidth]{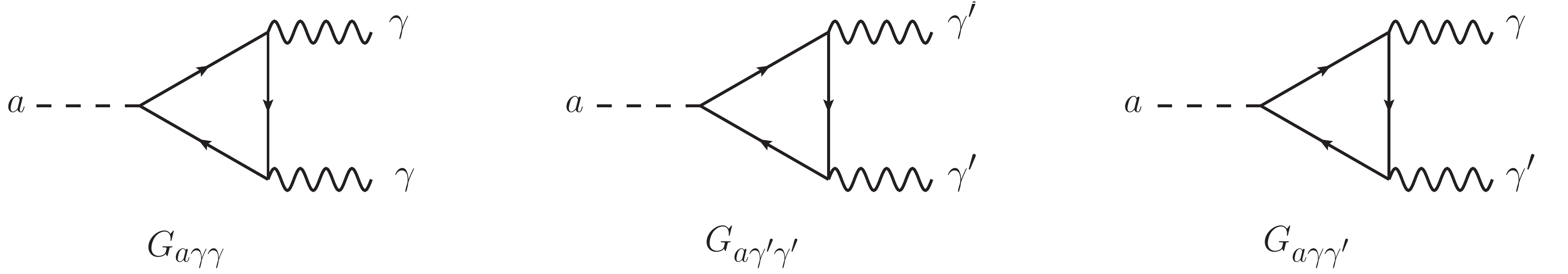}
\hfill
\caption{\label{fig:couplings} Extended axion couplings. The latter two are the new couplings from the dark axion portals.}
\end{figure}

In this Letter, we introduce and investigate a new portal that emerges when a dark photon and an axion coexist.
New vertices (axion--photon--dark photon, axion--dark photon--dark photon) materialize through the triangle diagrams\footnote{In the mirror symmetry models \cite{Berezhiani:2000gh,Ejlli:2016asd}, where the massless mirror photon is present, the axion-mirror photon-mirror photon coupling may exist.
In some sense, it corresponds to $G_{a\gamma'\gamma'}$, but $G_{a\gamma\gamma'}$ is absent in this kind of models.
Strictly speaking, the $G_{a\gamma'\gamma'}$ is not a traditional portal that connects the SM sector to the hidden sector.} (see Fig.~\ref{fig:couplings}).
Intriguingly, the new vertices are not just from a product of two individual portals; rather they are genuinely new couplings.
It is because of the new colored fermions that are charged under both the gauged $U(1)_\text {Dark}$ and the global $U(1)_{PQ}$.
Therefore, the dark photon can be attached to the triangle loop directly, not necessarily through the mixing of the vector portal.
Such new colored fermions are required to construct the KSVZ-type axion models \cite{Kim:1979if,Shifman:1979if}.
We shall call the new portal that connects the dark photon and the axion physics the ``dark axion" portal.

The dark photon and axion are widely studied light hypothetical particles with active experimental searches around the world.
The typical search schemes are based on the portal interaction of each particle, yet this might be incomplete in the presence of both particles.
The nature of the dark axion portal illustrates that the implications, including the best search schemes, could be beyond the native expectations.

This Letter will proceed as follows.
After we describe the dark axion portal terms, we introduce a simple model that realizes the new portal.
We provide one possible implication (among the many available) by presenting a new production mechanism for the dark photon DM in the early universe using the dark axion portal.

Although for definiteness we choose a QCD axion that addresses the strong $CP$ problem, many discussions here can be extended for the ALP.
Comprehensive studies including the ALP case will be given elsewhere \cite{future1}.

%%%%%%%%%%%%%%%%
{\em Dark axion portal.---}
%%%%%%%%%%%%%%%%
After a dark photon ($\gamma'$ or $Z'$) gets a small mass, the vector portal well below the electroweak scale is given by
\bea
{\cal L}_\text{vector portal} = \frac{\eps}{2}F_{\mu\nu}Z'^{\mu\nu} ,
\eea
where $F_{\mu\nu}$ and $Z'_{\mu\nu}$ are the field strengths of the photon and dark photon.
$\eps$ is the kinetic mixing parameter between the two $U(1)$ gauge symmetries \cite{Holdom:1985ag}, which is experimentally constrained to be very small ($\eps^2 \ll 1$) \cite{Essig:2013lka}.
The dark photon has been motivated from various dark matter related physics (such as the positron excess \cite{ArkaniHamed:2008qn}) and other physics (such as the $g_\mu - 2$ anomaly \cite{Gninenko:2001hx,Fayet:2007ua,Pospelov:2008zw}).

%%%%%%% FIGURE %%%%%%%
%%%%%%%%%%%%%%%%%%%

It should be mentioned that the full description before the symmetry breaking should be as a kinetic mixing between the $U(1)_Y$ hypercharge and the $U(1)_\text{Dark}$ \cite{fullDescription}.
However, it is sufficient to consider only the $U(1)_\text{QED}$ and the $U(1)_\text{Dark}$ mixing as long as $m_{\gamma'}^2 / m_Z^2 \ll 1$, which is why it is called the dark photon.
We will take this approach in this Letter, and postpone the generalization, including a full description, to a later work \cite{future1}.\footnote{Note that one can also take a gauge symmetry which has a nonzero charges for the SM particles as well as the kinetic mixing \cite{Lee:2016ief}.}

The axion portal is given by
\bea
{\cal L}_\text{axion portal} = \frac{G_{agg}}{4} a G_{\mu\nu}\tilde G^{\mu\nu} + \frac{G_{a\gamma\gamma}}{4} a F_{\mu\nu}\tilde F^{\mu\nu} + \cdots , ~~ \,
\eea
where $G_{\mu\nu}$ is the gluon field strength, tilde notation is for the dual, and $a$ is the axion.

The global Peccei-Quinn symmetry $U(1)_{PQ}$ was introduced to solve the strong $CP$  problem \cite{Peccei:1977hh,Peccei:1977ur}, which predicts a pseudo-Nambu-Goldstone boson, axion after the symmetry breaking \cite{Weinberg:1977ma,Wilczek:1977pj}.
In the invisible axion models \cite{Kim:1979if,Shifman:1979if,Dine:1981rt,Zhitnitsky:1980tq}, the PQ symmetry is broken spontaneously at an energy scale much higher than the electroweak scale.
The axion portal also applies to the ALP, which does not address the strong $CP$ problem.

In the presence of the dark photon, an axion can couple to two photons, two dark photons as well as a photon and a dark photon (see Fig.~\ref{fig:couplings}).
These interactions are given by the nonrenormalizable dark axion portal terms,
\bea
{\cal L}_\text{dark axion portal}= \frac{G_{a\gamma^\prime\gamma^\prime}}{4}  a Z'_{\mu\nu}\tilde Z'^{\mu\nu}
+\frac{G_{a\gamma\gamma^\prime}}{2} a F_{\mu\nu}\tilde Z'^{\mu\nu} . ~~~ \,
\eea
They are induced by the axion anomalous triangle couplings to (dark) photons as shown in Fig.~\ref{fig:couplings}.
Note the fermions in the triangle loop can have both $U(1)_{PQ}$ and $U(1)_\text{Dark}$ charges as well as the electromagnetic charges, and the dark photon can couple to them directly.

%%%%%%%%%%%%%%%%
{\em Dark KSVZ model.---}
%%%%%%%%%%%%%%%%
There is more than one way to realize the dark axion portal;  here we provide one of the simplest models based on the KSVZ type axion \cite{Kim:1979if,Shifman:1979if}, and we call it the dark KSVZ model.
(Other models, including the ALP case, will be studied elsewhere \cite{future1}.)

We take the $U(1)_\text{Dark}$ gauge symmetry and the $U(1)_{PQ}$ global symmetry.
We introduce only two Weyl fermions $\psi$ and $\psi^c$, which are colored singlets.
Note that our setup is free from the domain wall issue \cite{Sikivie:1982qv} since we have introduced only one flavor of heavy quarks.
They are vectorial under the gauge symmetries including the $U(1)_\text{Dark}$ so that the model is anomaly-free, yet chiral under the $U(1)_{PQ}$.
We also introduce two Higgs singlets $\Phi_{PQ}$ and $\Phi_D$ whose $CP$-odd components become an axion and the dark photon longitudinal mode.
Table~\ref{tab:charges} shows the new fields and their charge assignments in our model.
All SM fields have zero charges under the $U(1)_\text{Dark}$ and $U(1)_{PQ}$.
Yukawa terms for the exotic fermions are given by
\bea
{\cal L}_\psi = y_\psi \Phi_{PQ} \psi \psi^c + \text{H.c.},
\eea
which dictates $PQ_\Phi = - (PQ_\psi + PQ_{\psi^c})$.

As $\Phi_{PQ}$ develops a vacuum expectation value (VEV), the $U(1)_{PQ}$ is spontaneously broken at a scale $f_a$ (axion decay constant) given by $f_a^2 = PQ_\Phi^2 v_{PQ}^2 \label{eq:fa}$ with $\Phi_{PQ} = 1/\sqrt{2} (S_{PQ}+v_{PQ}) e^{i PQ_\Phi (a / f_a)}$.
It gives a mass to the exotic fermions $m_\psi = (y_\psi / \sqrt{2}) v_{PQ}$.
After the QCD phase transition, the axion mass is given by
$m_a \simeq [\sqrt{z}/(1+z)] (f_\pi / f_a) m_\pi$, where $z=m_u/m_d\simeq0.56$, and $m_\pi\simeq 135 ~\mev$ and $f_\pi\simeq 92 ~\mev$ are the mass and the decay constant of the pion, respectively.
For the spontaneous breaking of $U(1)_\text{Dark}$, as $\Phi_D$ develops a VEV, $\langle\Phi_D\rangle=v_D/\sqrt{2}$, the dark photon acquires a mass given by $m_{\gamma'}^2 = e'^2 D_{\Phi_{D}}^2 v_D^2$.

The axion-gluon-gluon coupling is given by
\bea
G_{agg} = \frac{g_S^2}{8\pi^2} \frac{PQ_\Phi}{f_a} .
\eea
We derive the axion portal and the dark axion portal terms below the QCD scale in the dark KSVZ model at the leading order in $\eps$ as follows.
\bea
G_{a\gamma\gamma} &=& \frac{e^2}{8\pi^2} \frac{PQ_\Phi}{f_a} \Big[ 2 N_C Q_\psi^2 - \frac{2}{3} \frac{4+z}{1+z} \Big] , \label{eq:Gagg} \\
G_{a\gamma\gamma'} &\simeq& \frac{e e'}{8\pi^2} \frac{PQ_\Phi}{f_a} \big[ 2 N_C D_\psi Q_\psi \big] + \eps G_{a\gamma\gamma} , \\
G_{a\gamma'\gamma'} &\simeq& \frac{e'^2}{8\pi^2} \frac{PQ_\Phi}{f_a} \big[ 2 N_C D_\psi^2 \big] + 2 \eps G_{a\gamma\gamma'} .
\eea
$N_C = 3$ is the color factor, $g_S$ is the $SU(3)_C$ coupling, and $e'$ is the $U(1)_\text{Dark}$ coupling, which can be as sizable as the SM gauge couplings.
Note the $PQ_\Phi$ dependence is not real as it is canceled by the same factor in $f_a$.

These expressions clearly show that the dark axion portal couplings are not just from the product of two individual portals (e.g., $\eps G_{a\gamma\gamma}$, which is greatly suppressed because $\eps^2 \ll 1$); rather, they are new relatively large couplings originating from the dark gauge symmetry.
As a matter of fact, the new couplings $G_{a\gamma\gamma'}$, $G_{a\gamma'\gamma'}$ may be still sizable even when the vector portal and axion portal vanish ($\eps \simeq 0$, $G_{a\gamma\gamma} \simeq 0$).
The $G_{a\gamma\gamma'}$ would vanish if the new colored fermions do not carry electric charges ($Q_\psi = 0$), except for the part induced by the vector portal ($\eps$).%\footnote{These couplings would change if there are more fermions, e.g. some exotic leptons, that can be in the triangles of Fig.~\ref{fig:couplings}.}

Now, we discuss the effect of the exotic colored fermions on $\eps$ by the loop.
The loop-induced contribution is given by $\eps_\text{induced} = (1 / 6 \pi^2) N_C (e Q_\psi e' D_\psi) \log (\Lambda / m_\psi)$ in which $\Lambda$ is the scale where $\eps_\text{induced} = 0$.
If we take $\Lambda$ as a typical grand unified theory scale and $m_\psi \sim f_a$, we get $\eps_\text{induced}$ $\sim$ ${\cal O} (10^{-3} - 10^{-2})$ for $e' \sim e$ and $Q_\psi \sim D_\psi \sim 1$.
This is on its own inconsistent with the experimental constraints for the keV--MeV scale dark photon \cite{Essig:2013lka}.
This can be addressed either (i) by assuming a cancellation between the loop-induced contribution and the short-distance (or UV) contribution to $\eps$ (in spite of fine-tuning), or (ii) by introducing more particles that couple to $\gamma$ and $\gamma'$ in a similar way to Refs.~\cite{Holdom:1985ag,Davoudiasl:2012ig} (although this means increasing the model complexity).

Stability of the exotic colored fermion ($\psi$, $\psi^c$) is a generic problem of the KSVZ model.
This unwanted challenge can be avoided in our model by an appropriate charge assignment for these particles.
For example, for the choice of $PQ_\psi =0$, $Q_\psi=-1/3$ and $D_\psi = D_\Phi$, the $\Phi_D^\dagger \psi D^c$ term is allowed.
Then the exotic fermion can decay into the SM down-type quark and the SM singlet scalar $S_D$, which can subsequently decay into the dark photons.

The Higgs portal can be also introduced in this model because of the existence of $\Phi_{PQ}$ and $\Phi_D$.
%%%%%%%%%%%%% TABLE %%%%%%%%%%%%%%
\begin{table}[tb]
\begin{tabular}{c|ccc|cc}
\hline
~Field~ & $SU(3)_C$ & $SU(2)_L$ & $U(1)_Y$ & $U(1)_\text{Dark}$ & $U(1)_{PQ}$ \\
\hline
%$Q$    & $3$         & $2$ & $1/6$ & $0$ & $0$ \\
%$U^c$ & $\bar 3$ & $1$ & $-2/3$ & $0$ & $0$ \\
%$D^c$ & $\bar 3$ & $1$ & $1/3$ & $0$ & $0$ \\
%$L$     & $1$        & $2$ & $-1/2$ & $0$ & $0$ \\
%$E^c$ & $1$ & $1$ & $1$ & $0$ & $0$ \\
%\hline
$\psi$    &  $3$ & $1$ & $Q_\psi$  & $D_\psi$   & $PQ_\psi$ \\
$\psi^c$ & $\bar 3$ & $1$ & $-Q_\psi$ & $-D_{\psi}$ & $PQ_{\psi^c}$ \\
\hline
$\Phi_{PQ}$    & $1$ & $1$ & $0$ & $0$ & $PQ_\Phi$ \\
$\Phi_D$    & $1$ & $1$ & $0$ & $D_\Phi$ & $0$ \\
\hline
\end{tabular}
\caption{New fields and charge assignments in our model. $Q_\psi$ is the electromagnetic charge of the exotic fermion $\psi$.}
\label{tab:charges}
\end{table}
%%%%%%%%%%%%%%%%%%%%%%%%%%%%%%%%%%

%%%%%%%%%%%%%%%%
{\em Decays.---}
%%%%%%%%%%%%%%%%
The dark photon decay widths are
\bea
\Gamma(\gamma' \to e^+ e^-) &=& \frac{\eps^2 e^2}{12\pi} m_{\gamma'} \Big[1 - \frac{4 m_e^2}{m_{\gamma'}^2} \Big]^{1/2} , \\
\Gamma (\gamma' \to \gamma a) &=& \frac{G_{a \gamma \gamma'}^2}{96 \pi} m_{\gamma'}^3 \Big[ 1 - \frac{m_a^2}{m_{\gamma'}^2} \Big]^3 .\label{eq:gPtoga}
\eea
This can be compared to the $\Gamma(\gamma' \to 3 \gamma) \approx (5 \times 10^{-8}) \eps^2 (e^2 / 4\pi^2)^4 (m_{\gamma'}^9 / m_e^8)$ \cite{Pospelov:2008jk}, which would be the dominant decay mode for a sub-MeV dark photon if there is no dark axion portal decay of $\gamma' \to \gamma a$.

The axion decay widths are
\bea
\Gamma (a \to \gamma \gamma) &=& \frac{G_{a\gamma\gamma}^2}{64\pi} m_a^3 , \\
\Gamma (a \to \gamma \gamma') &=& \frac{G_{a\gamma\gamma'}^2}{32\pi} m_a^3 \Big[1-\frac{m_{\gamma'}^2}{m_a^2}\Big]^3 , \\
\Gamma (a \to \gamma' \gamma') &=& \frac{G_{a\gamma'\gamma'}^2}{64\pi} m_a^3 \Big[1-\frac{4m_{\gamma'}^2}{m_a^2}\Big]^{3/2} .
\eea
The $a$-$e^+$-$e^-$ coupling is two-loop suppressed in our model as it is a hadronic axion (of KSVZ type).

%%%%%%%%%%%%% FIGURE %%%%%%%%%%%%%
\begin{center}
\begin{figure}[t]
\includegraphics[width=0.3\textwidth]{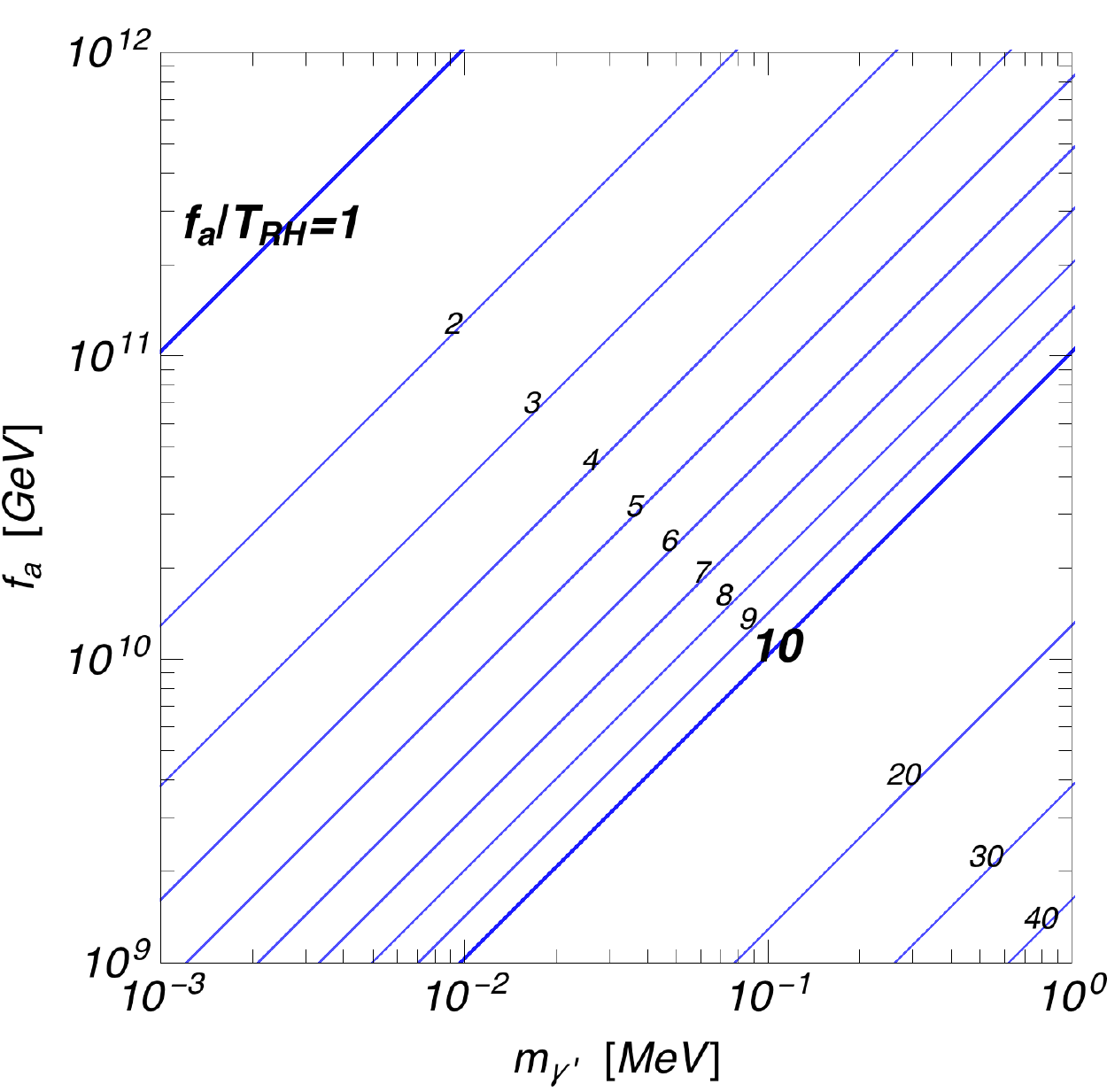}
\caption{The blue lines show $\Omega_{\gamma'} h^2 = 0.12$ for the given $f_a / T_{\rm RH}$ values.
We choose $e'=0.1$, $D_\psi=3$, $Q_\psi= 0$ with $g_* = 100$.
For low $f_a$, the axion alone cannot explain the observed relic density, yet it can be accounted for with the dark photon.
}
\label{fig:fig_dm}
\end{figure}
\end{center}
%%%%%%%%%%%%%%%%%%%%%%%%%%%%%%%

%%%%%%%%%%%%%%%%
\vspace{-8mm}
{\em Implications in the cosmology.---}
%%%%%%%%%%%%%%%%
There may be many applications of the dark axion portal, and here we illustrate one in the cosmology with the dark KSVZ model.
We present a novel production mechanism of the dark photon DM candidate in the early universe using the dark axion portal, which can also address an issue of the axion DM relic density.
Known production mechanisms of the dark photon are very rare \cite{Redondo:2008ec,Nelson:2011sf}, although physics of the dark photon DM physics is widely investigated \cite{Arias:2012az}.

In the QCD axion model, the viable window of the $f_a$ is given by $10^9 \lesssim f_a \lesssim 10^{12}~{\rm GeV}$.
The lower bound is from the SN1987A data, and the upper bound is from the relic density constraint ($\Omega_{\rm DM} h^2 = 0.12$), which is obtained from the misalignment mechanism \cite{Kim:2008hd}.
Below $10^{11}~{\rm GeV}$, the required relic density may not be explained by the axion DM alone.
We take the dark photon as another DM candidate produced by the freeze-in mechanism \cite{McDonald:2001vt,Hall:2009bx} through the dark axion portal.
This additional dark matter can compensate for the required relic density.

For definiteness, we take $Q_\psi=0$ and $\eps \simeq 0$ so that the decay channel $\gamma'\to a\gamma$ is forbidden and the dark photon is stable\footnote{Exotic colored fermions might be stable for this case. As we will see, however, their production would be little since the reheating temperate is lower than their mass scale, $f_a$, in the relevant parameter space.}\,\footnote{The scale dependence of $\eps$ due to the gauge coupling running \cite{delAguila:1988jz,Davoudiasl:2015hxa} does not apply here.}, and we consider the dark photon to be of roughly keV scale.

The main process to connect the SM sector with the dark sector is $gg \leftrightarrow \gamma'\gamma'$ mediated by the axion.
Because of a large $f_a$, this reaction is feeble and cannot bring $\gamma'$ into the thermal bath.
Then, the freeze-in mechanism ($gg \to \gamma'\gamma'$) can work to produce the non-thermal $\gamma'$.

The Boltzmann equation for $\gamma'$ is given by
\begin{eqnarray}
	%\frac{dn_{\gamma'}}{dt}+3Hn_{\gamma'}=
	-SHT\frac{dY_{\gamma'}}{dT}=\gamma[n_{\gamma'}],
	\qquad
	Y_{\gamma'}\equiv n_{\gamma'}/S,
\end{eqnarray}
where $\gamma[n_{\gamma'}]$ denotes the collision term, and we use $S=(2\pi^2/45)g_{*s}T^3$ (entropy density) and $H^2=(\pi^2/90)g_{*\rho}T^4/M_{\rm Pl}^2$ (Hubble parameter) with $M_{\rm Pl} \simeq 2.4\times10^{18}$ GeV.
In the following discussion, we take $g_{*s}=g_{*\rho}\equiv g_*$ as a constant value.

We take only the dominant gluon contribution, and the annihilation cross section is approximately given by $\sigma v \simeq 4 G_{agg}^2G_{a\gamma'\gamma'}^2 s$.
We then obtain
\begin{eqnarray}
	\gamma[n_{\gamma'}] &\simeq& \frac{48}{\pi^4}G_{agg}^2G_{a\gamma'\gamma'}^2T^8 , \label{eq:coll}
\end{eqnarray}
where we neglect $m_{\gamma'}$; thus, by integrating $T$ from $T_{\rm RH}$ to 0, we end up with $Y_{\gamma'}^0\equiv Y_{\gamma'}(T=0)$ as
\begin{eqnarray}
	Y_{\gamma'}^0
	&\simeq&
	\frac{1080\sqrt{10}}{\pi^7g_*^{3/2}}G_{agg}^2G_{a\gamma'\gamma'}^2M_{\rm Pl}T_{\rm RH}^3 ,
\end{eqnarray}
where $T_{\rm RH}$ is the reheating temperature.
Then, we obtain 
\begin{eqnarray}
	&&\Omega_{\gamma'} h^2 = S_0 m_{\gamma'}Y_{\gamma'}^0(\rho_c/h^2)^{-1}\nonumber\\ 
	&\simeq&
	0.1\times g_D^4
	\Big[	\frac{10^2}{g_*} \Big]^\frac{3}{2}
	\Big[	\frac{m_{\gamma'}}{10 \, {\rm keV}} \Big]
	\Big[	\frac{5 \, T_{\rm RH}}{f_a} \Big]^3
	\Big[	\frac{10^{10}}{f_a / {\rm GeV}} \Big] , ~
\end{eqnarray}
where $\rho_c=1.05368\times10^{-5}h^2~{\rm GeV}~{\rm cm}^{-3}$ and $S_0=2889.2~{\rm cm}^{-3}$ are the critical density and the entropy density at the present time.
We define $g_D\equiv(e'D_\psi/0.3)$.
We use $f_a = v_{PQ}$ without loss of generality.

The viable parameter region is shown in Fig.~\ref{fig:fig_dm} for a choice of parameter values.
One can take low $T_{\rm RH}$ to reduce the $\gamma'$ relic density, and can always find the total relic density of the $a$ and $\gamma'$ system that can explain the observed relic density.

It should be noted that there is an upper limit on $T_{\rm RH}$ for the freeze-in DM production to work, because at $T=T_{\rm RH}$ the DM annihilation takes place most frequently and it might be thermalized if $T_{\rm RH}$ is sufficiently high.
This upper bound can be obtained by imposing $r_{\gamma'} < H$ at $T=T_{\rm RH}$, where the reaction rate is $r_{\gamma'}\equiv\gamma[n_{\gamma'}]/n_{\gamma'}^{\rm eq}$ with $n_{\gamma'}^{\rm eq}\simeq [3\zeta(3)/\pi^2] T^3$.
Then, the upper bound becomes
\begin{eqnarray}
	T_{\rm RH}\lesssim (10^{10}~{\rm GeV})
	\times g_D^{-\frac{4}{3}}
	\Big[	\frac{g_*}{10^2} \Big]^\frac{1}{6}
	\Big[	\frac{f_a/{\rm GeV}}{10^{10}} \Big]^{\frac{4}{3}}
\end{eqnarray}
above which $\gamma'$ may be thermalized, leading to overproduction.
This may still be consistent with the condition $T_{\rm RH}\gtrsim10^9~{\rm GeV}$, which is required in the thermal leptogenesis \cite{Davidson:2002qv}.

Finally, let us comment on the case of $Q_\psi\neq0$.
The dark photon would decay ($\gamma'\to a\gamma$) through $G_{a\gamma\gamma'}$ and bring a constraint with its lifetime.
We would also need to take into account another $\gamma'$ production process similar to the Primakoff effect \cite{Bolz:2000fu}.
With the $G_{a\gamma\gamma'}$ coupling, a keV--MeV scale dark photon can be constrained by the astrophysical objects such as the globular cluster \cite{Redondo:2008ec} and the supernova \cite{Chang:2016ntp} in a similar way the axion is constrained \cite{Kolb:1990vq,Raffelt:1996wa}.
Emission of the axion produced by a bremsstrahlung process using the axion coupling to the electron or nucleon, especially with the SN1987A data, gives the strongest bound $f_a \gsim 10^9$ GeV, and the bound through the axion coupling to the photon is much weaker.
Therefore, the $a$ and $\gamma'$ system in this range would be safe from these constraints even for the $G_{a\gamma\gamma'} \sim G_{a\gamma\gamma}$ case.\footnote{Constraint on $G_{a\gamma\gamma'}$ in general can be obtained by, for instance, a plasmon decay ($\gamma^* \to \gamma' a$), which calls for further study.}

%%%%%%%%%%%%%%%%%%%%%%%%%%%%
{\em Summary and outlook.---}
%%%%%%%%%%%%%%%%%%%%%%%%%%%%
Many studies about the hidden sector have been performed in the framework of the portals, including the vector portal and the axion portal, which are useful for understanding possible couplings and determining the best search methods for the dark photon and axion (or ALP), respectively.
They were extensively studied and experimentally searched.

We investigated the physics and phenomenological consequences when both new particles exist together.
A new portal connecting the dark photon and axion ($G_{a\gamma\gamma'}$, $G_{a\gamma'\gamma'}$) emerges.
The new couplings can be much stronger than the simple product of the vector portal and axion portal couplings.
Typical search schemes could be invalidated in some cases, and new search schemes based on this dark axion portal could be more important.

We exhibited the dark axion portal by constructing a new model based on the KSVZ axion model.
Because of the nature of the QCD axion, which has an extremely tiny mass and feeble couplings, the relevant implications are in the cosmology and/or astroparticle physics area.
We illustrated the usefulness of the new portal with an example of the relic density issue of the axion dark matter using the $gg - a - \gamma' \gamma'$ channel.
The dark axion portal provides a novel mechanism for the dark photon production, and a sufficiently light dark photon can be a dark matter candidate that can compensate for the deficit of the relic density of the axion dark matter when the axion decay constant is low.
Various channels of production and decay for the axion or dark photon are allowed through the dark axion portal; the traditional implications should be revisited in this new picture.

The dark axion portal is not specific to the QCD axion; it can also be applied to the ALP whose mass and couplings can be large enough so that many laboratory experiments, such as rare meson decays or beam dump experiments, are relevant.
Various model buildings realizing the dark axion portal and rich phenomenology are warranted.\footnote{Recently, there have been some efforts to combine the axion and dark photon together \cite{Ejlli:2016asd,relaxion}.}
As all other portals have done in the past, the introduction of a new portal will open great opportunities for many subsequent studies.

%---------------------------------------------------------
\begin{acknowledgments}
{\em Acknowledgments.---}
This work was supported by IBS (Project Code IBS-R018-D1).
H.L. thanks J.Y. Lee and C.S. Kim for useful discussions at the start of this project during a visit to Yonsei University.
Especially, some part of the formalism used in this paper started to be developed with J.Y. Lee in the early stage of the project.
H.L. also thanks S.H. Im for the conversation on the topic.
\end{acknowledgments}
%---------------------------------------------------------

%---------------------------------------------------------

\end{document}